**Quantum simulation of Kibble-Zurek mechanism with a semiconductor electron charge qubit**

Cheng Zhou[1,*], Li Wang[1,*], Tao Tu[1], Hai-Ou Li[1], Guang-Can Guo[1], Hong-Wen Jiang[2] and Guo-Ping Guo[1]

[1] Key Laboratory of Quantum Information, University of Science and Technology of China, Chinese Academy of Sciences, Hefei 230026, People's Republic of China

[2] Department of Physics and Astronomy, University of California at Los Angeles, California 90095, USA.

[*] These authors contributed equally to this work. Correspondence and requests for materials should be addressed to T. T. (tutao@ustc.edu.cn) or to G. P. G. (gpguo@ustc.edu.cn).

**The Kibble-Zurek mechanism provides a description of the topological structure occurring in the symmetry breaking phase transitions, which may manifest as the cosmological strings in the early universe or vortex lines in the superfulid. A particularly intriguing analogy between Kibble-Zurek mechanism and a text book quantum phenomenon, Landau-Zener transition has been discovered, but is difficult to observe up to now. In recent years, there has been broad interest in quantum simulations using different well-controlled physical setups, in which full tunability allows access to unexplored parameter regimes. Here we**

**demonstrate a proof-of-principle quantum simulation of Kibble-Zurek mechanism using a single electron charge qubit in double quantum dot, set to behave as Landau-Zener dynamics. We measure the qubit states as a function of driven pulse velocity and successfully reproduce Kibble-Zurek like dependence of topological defect density on the quench rate. The high-level controllability of semiconductor two-level system make it a platform to test the key elements of topological defect formation process and shed a new insight on the aspect of non-equilibrium phase transitions.**

The Kibble-Zurek mechanism[1-2] predicts the production of topological structures such as strings, vortices, or domain walls in the symmetry breaking phase transitions. This prediction is first discussed in cosmological phase transition which occurs as the early universe expands and cools[1,3]. Then similar process has widely applied to the phase transitions which take place in the laboratory: superfluid helium in an annulus, liquid crystal or superconducting loop[2,4,5]. The Landau-Zener transition[6,7] has played a prominent role in the quantum phenomena, that occurs when a two-level system sweeping through its anti-crossing point. There has been increasing interest to observe the Landau-Zener process in both natural atoms and artificial semiconductor quantum dots and superconducting junctions[8-11]. In 2005, Damski presented that the Landau-Zener transition is a simplest quantum model exhibiting the key physics of Kibble-Zurek mechanism[12]. This unexpected insight has been an extensive subject of discussion in the past years[13-16].

The analogue quantum simulation aims to simulate a quantum system using a controllable artificial system that has the same mathematical model[17]. It provides an efficient way to simulate quantum systems over a wide range of parameter regimes which can not be accessed on a classical computer or experimentally. Various quantum simulators have been implemented on atomic quantum gases[18], ensembles of trapped ions[19], photonic systems[20] and superconducting circuits[21]. Electron charge[22,23] or spin qubits[24] in semiconductor quantum dots are particularly interesting for the candidate of quantum simulation, as they allow high level of control of various experimental parameters. Here we demonstrate an analogue quantum simulation scheme of Kibble-Zurek mechanism using a single electron charge qubit in a double quantum dot.

**Results**

**Analogy between Kibble-Zurek mechanism and Landau-Zener transition**

Kibble-Zurek mechanism gives a natural description of topological defects formation in the symmetry breaking phase transitions occurring from cosmology to condensed matter[1-5]. As shown in Figure 1a, consider a system traversing a second order phase transition during a quench time. Here we define the distance from the critical point as $\varepsilon_{KZ}$. The ratio of time $t$ and $\varepsilon_{KZ}$ yields the quench time $\tau_Q = t / \varepsilon_{KZ}$. When the system approaches the critical point from the high symmetry phase, the order parameters of the system such as relaxation time $\tau = \tau_0 / \varepsilon_{KZ}$ will become divergence which implies its dynamics becomes increasing slow. In particular, when the relaxation time are larger than the quench time, also called the freeze-out time scale $\hat{t}_{KZ}$ [3-5], the new broken symmetry states are in effect immobilized, and topological defects form. Then the density of the defects $\rho_d$ that is the average size of the topological defects, is given by the correlation length at the freeze-out time, which is the key prediction of the Kibble-Zurek mechanism[1-5]. According to the scale of relaxation time and quench time, the non-equilibrium phase transition process can be distinguished into two regions in the neighborhood of the critical point: adiabatic freezing and non-adiabatic freezing.

A transition between two energy levels at the anti-crossing point is known as Landau-Zener transition[6,7]. To be more specific, in schematic Figure 1b, we consider a two-level system at a sweeping rate $\nu$. Here we denote the level detuning as $\varepsilon_{LZ}$, the ground and excited states as $|0\rangle$ and $|1\rangle$, respectively. When the system approaches to the anti-crossing point, there is a characteristic boundary $\hat{t}_{LZ}$ [8] where the energy gap $E_\Delta$ between the up and down levels is minimized in comparison to the sweeping rate, to allow for a sudden transition between two levels. Thus it is also intuitive to use a two-stage picture[8]: the system will evolve almost adiabatically along the ground state far form the anti-crossing point and have a significant non-adiabatic transition probability $P_{LZ}$ to the excited state in the vicinity of the anti-crossing point.

Compared with the two phenomena as illustrated in Figure 1a and 1b, Damski[12] pointed out that the two-level system undergoes the Landau-Zener transition from ground state to the excited state, in the similar way as quenched phase transition from a defected-free phase to a defected one. The characterization quantities in the Kibble-Zurek process such as quench time $\tau_Q$, relaxation time $\tau$, and resulted defect density $\rho_d$ can be mapped onto the inverse of the sweeping rate $v$, the inverse of the energy gap $E_\Delta$, and transition probability $P_{LZ}$ in the Landau-Zener picture, respectively[12]. Analog between the two phenomena allow for a shortcut towards quantum simulations[17-21]. Here we would like to simulate a given process (Kibble-Zurek mechanism) by a different one (Landau-Zener transition) containing all the essential features of the original problem. Direct quantum simulation of the topological defect formation in a non-equilibrium second order phase transition (the main result of Kibble-Zurek mechanism) can be implemented by controllable evolution of the initial state into the corresponding final state of interest in a Landau-Zener transition.

**Landau-Zener dynamics of a electron charge qubit in double quantum dot**

The quantum simulator system needs to be in a well-control manner. For the simulation, we use a single electron charge qubit trapped in a double quantum dot. The quantum bit Hamiltonian in the basis of $|L\rangle$ and $|R\rangle$[22,23],

$$H_{LZ} = \frac{1}{2}\varepsilon_{LZ}\sigma_Z + \Delta\sigma_X, \qquad (1)$$

consists of two terms. The first term denotes the energy level difference of the states $|L\rangle$ and $|R\rangle$, and the second term represents the interaction between the two states, depending on the tunnel coupling of two dots. Here we denote electron in the left and right dot as the states $|L\rangle$ and $|R\rangle$, respectively. In the simulation, the first term in equation (1), can be coupled and tuned with a time-dependent electrical field (details of sample structures, experimental techniques and characterization of the qubit are given in the Methods section).

Our scheme to probe Landau-Zener transition of the qubit using a double passage process is shown in Figure 2. Firstly we initialize the quantum simulator to the ground state $|0\rangle$ at a positive detuning $\varepsilon_{LZ0}$. We drive the qubit in a superposition state of $|0\rangle$ and $|1\rangle$ via a linear raising pulse through the anti-crossing region. Note that the Landau-Zener transition has a probability $P_{LZ} = \exp(-2\pi\Delta^2/\hbar v)$. Then we apply a linear decreasing pulse to take the qubit back to pass the anti-crossing point twice. This double-passage process will give the final state of the qubit as a function of the non-adiabatic level transition $P_{LZ}$ and accumulated phase $\phi$ during the adiabatic evolution[8,10]:

$$P_{|1\rangle} = 2P_{LZ}(1-2P_{LZ})[1+\cos\phi]. \qquad (2)$$

The probability of finding the qubit in the excited state $|1\rangle$ is measured using the nearby quantum point contact charge sensor. Consequently, the charge state probability $P_{|1\rangle}$ oscillates as a function of various control parameters, known as the Landau-Zener-Stückelberg interference[8,10].

**Quantum simulation of Kibble-Zurek prediction using Landau-Zener transition**

To study the quenched second order phase transition process such as Kibble-Zurek predicted topological defects formation, it is necessary to simulate the density of defects $\rho_d$ as a function of the quench rate $\tau_Q/\tau_0$. In more specific, the key prediction of Kibble-Zurek mechanism can be formulated as[12]:

$$\rho_{\mathrm{d}} = 2 / P(x_\alpha), \qquad (3)$$

where $x_\alpha = \alpha \tau_{\mathrm{Q}} / \tau_0$, $P(x_\alpha) = x_\alpha^2 + x_\alpha \sqrt{x_\alpha^2 + 4} + 2$ and $\rho_{\mathrm{d}}$ is normalized to unity.

It has been noted theoretically[12] that two quantities $\rho_{\mathrm{d}}$ and $\tau_{\mathrm{Q}} / \tau_0$ in the Kibble-Zurek mechanism are equivalent to the Landau-Zener transition probability $P_{\mathrm{LZ}}$ and $4\Delta^2 / \hbar v$, respectively. We simulate the equation (3) by continuously adjusting the amplitude of applied electrical pulse on the qubit while its time width fixed or inverse (see the Methods section for details). Precise control of the driven pulse enables sweeping rate $v$ over a wide range, ensuring that Landau-Zener transition probability $P_{\mathrm{LZ}}$ can be investigated in the crossover from adiabatic to non-adiabatic dynamics. In the experiment, $\Delta$ is fixed and determined as 10.3 μeV.

Figure 3 shows the recorded charge state occupation $P_{|1\rangle}$ as a function of both the energy level detuning $\varepsilon_{\mathrm{LZ0}}$ and the sweeping rate $v$. In the absence of decoherence, the visibility of the Landau-Zener-Stückelberg interference pattern is given by $V = 4P_{\mathrm{LZ}}(1 - 2P_{\mathrm{LZ}})$[25], meaning that $P_{\mathrm{LZ}}$ can be determined by monitoring the information about $P_{|1\rangle}$. In practice, we obtained the value of $P_{\mathrm{LZ}}$ from the raw data in the form[25,26] that accounts for the general effects including the charge noise in the surround environment, the imperfect fidelity of the driven pulse, and the inefficiency of the measurement.

In Figure 4, we plot the simulated $\rho_{\mathrm{d}}$ as a function of $\tau_{\mathrm{Q}} / \tau_0$. The experimental data is in reasonable agreement with the theoretical model based on equation (3), which is shown as solid line. From these data, it can be seen that as the quench rate decreases, whereas the density of the topological defects grows, show the crossover from the near equilibrium to non-equilibrium limits. Hence the data confirm the Kibble-Zurek mechanism expected theoretically.

**Discussion**

In summary, we have implemented a proof-of-principle quantum simulation of a tunable non-equilibrium second order phase transition. We have demonstrated that the simulated dynamics for the topological effect formation shows a function of quench rate, one of the natural features of the Kibble-Zurek mechanism. Our experiment serves a first step to more complex quantum simulations. A route for the near future will be to move towards the quantum simulation of dynamics that are difficult to observe in the real systems such as superfluid[27]. Furthermore, the mapping between Kibble-Zurek mechanism and quantum dynamics would pave the way to address a broad range of fundamental issues in quantum phase transitions occurring in various condensed matter physics, such as one-dimensional quantum Ising model in a transverse field[16]. Our scheme does not use a sequence of quantum gates as digital quantum simulation[17-21] and the desired outcome does not be affected drastically by decoherence, thus scaling to a large number of quantum dot qubits might be simpler.

**Methods**

Our wafer is grown by molecular-beam epitaxy and a thin layer of electrons called two-dimensional electron gas can be formed at the interface between AlGaAs layer and GaAs layer. Using the standard Hall resistance data and Shubnikov-de Hass oscillations in the longitudinal resistance, the density and mobility of the two-dimensional electron gas are determined as $3.2\times10^{11}$ cm$^{-2}$ and $1.5\times10^{5}$ cm$^{2}$/Vs, respectively. Figure 2 shows a top-down scanning electron micrograph of one of our processed double quantum dot devices. The Ti/Au electrodes are fabricated on the top surface of the wafer, among them, the L2, L3, R2, R3, M, and T gates are used to confine the electrons in the double quantum dot, the M and T gates can also adjust the tunnel coupling between the two dots. The left and right quantum point contacts, which are used as nearby charge detectors, are formed by the L1 and L2, R1 and R2 gates, respectively.

An Agilent 81134A pulse generator is used to implement control pulse with designed amplitude and time width onto the L3 gate of the double quantum dot. In the experiment, we use a low-pass filter to change the pulse profile further. The energy level velocity $v$ is evaluated as the rate of the changing energy level spacing involved in the sweeping[10,28]. The double quantum dot charge configurations are measured via standard charge detection techniques[22,29]. Resulting from the strong capacitive coupling between double quantum dot and nearby detector, the changes in the conductance $G_{QPC}$ through the quantum point contact, are sensitive to the changes in the double quantum dot charge states. Here we use the right quantum point contact as the sensor. In practice, we usually apply a small modulation to the gate L1 and use lock-in method to measure the quantum point contact differential transconductance to enhance the signal sensitivity[29,30]. All measurements are performed in an Oxford

Triton dilution refrigerator with a base temperature of 30 mK.

quantum dot. *Nature* **481**, 344-347 (2012).

**Acknowledgements** We thank Dr. Gang Cao for helpful discussion. This work was supported by the National Fundamental Research Program (Grant No. 2011CBA00200), and National Natural Science Foundation (Grant Nos. 11274294, 11074243, 10934006, 11222438, 91121014).

**Author Contributions** T. T., L. W. and G. C. G. designed the experiment, provided theoretical support and analyzed the data. C. Z., H. O. L. and H. W. J. fabricated the samples and performed the measurements. G. P. G. supervised the project.

**Competing financial interests** The authors declare no competing financial interests.

**Figure Legends**

**Figure 1 | Analogy between Kibble-Zurek mechanism and Landau-Zener transition.** (a) Top: Schematic diagram of quenched symmetry breaking phase transition, resulting in the formation of topological defects $\rho_d$. Down: We can define two regions: Within (outside) the time interval $[-\hat{t}_{KZ}, \hat{t}_{KZ}]$, in which the relaxation time $\tau$ is longer (smaller) than the quench time $\tau_Q$, so that the phase configuration will be fixed (mobile). (b) Top: Illustration of the two-level system dynamics for the time-dependence sweeping. Down: We can also distinguish two regions: Near (far from) the anti-crossing point, in which the inverse of the energy gap $E_\Delta$ is smaller

(larger) than the sweeping rate $v$, so that there is a non-adiabatic transition $P_{\text{LZ}}$ between the two states (adiabatic evolution).

**Figure 2 | Experimental scheme to realize Landau-Zener transition of single electron charge qubit in double quantum dot.** Top: Scanning electron micrograph of the double quantum dot device and the quantum point contact charge sensor. The color plot of the electron represents the occupation states during the pulse sweeping. The pulse protocol is also indicated. Down: a controlled double passage Landau-Zener transition process is drawn.

**Figure 3 | Observation of the controllable Landau-Zener dynamics.** The occupation of the qubit in the excited state $|1\rangle$ as a function of the energy position $\varepsilon_{\text{LZ0}}$ and the sweeping rate $v$, revealing a double passage Landau-Zener transition process.

**Figure 4 | Quantum simulation of Kibble-Zurek like behavior of defect density as a function of quench rate.** The dots represents the normalized to unity density of the defects $\rho_d$ as a function of quench rate $\tau_Q / \tau_0$. The solid curve is the theoretical prediction based on equation (3). For the illustration of the analogy, the correspoding quantities in the Landau-Zener transition such as $P_{\text{LZ}}$ and $4\Delta^2 / \hbar v$ are also given in the brackets.

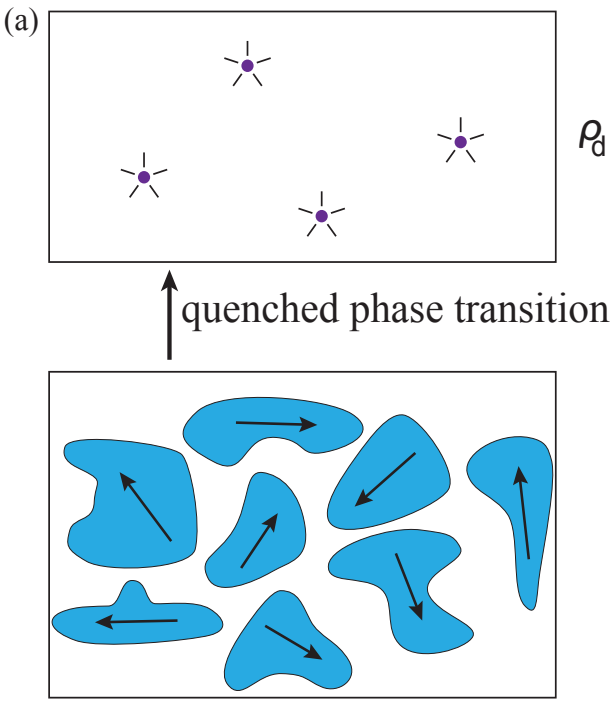
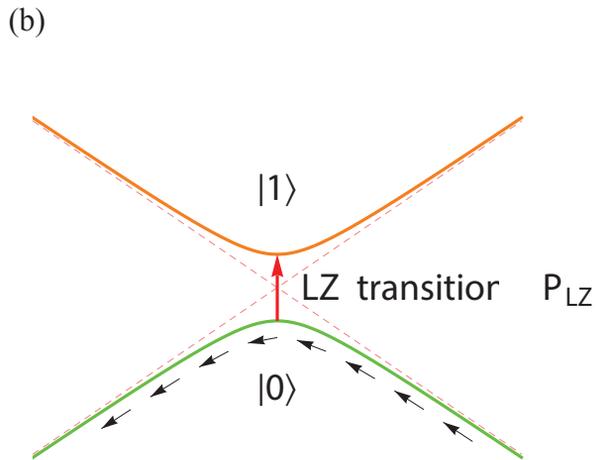
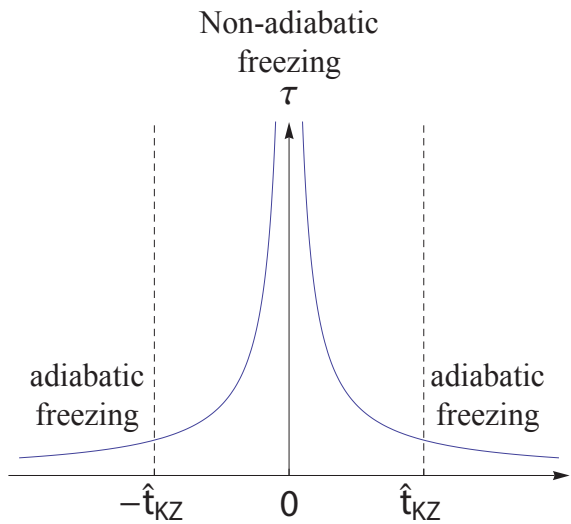
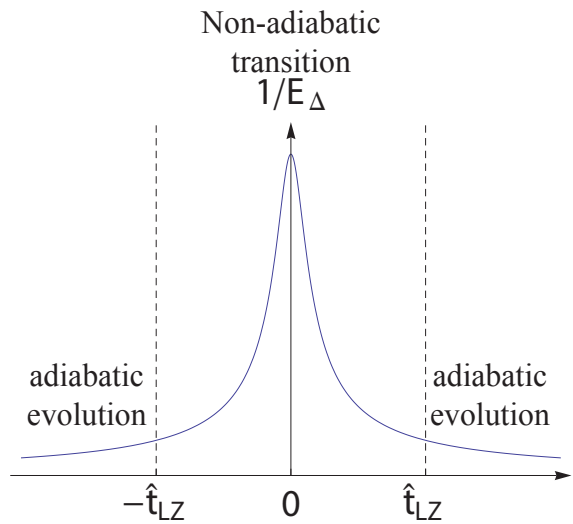

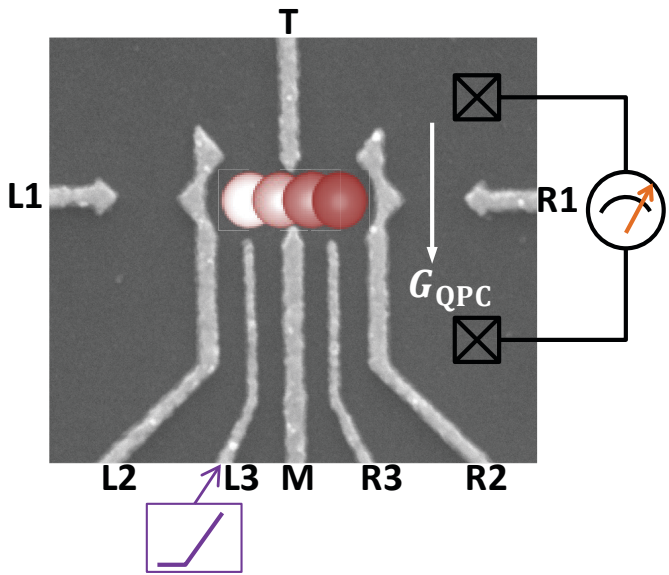 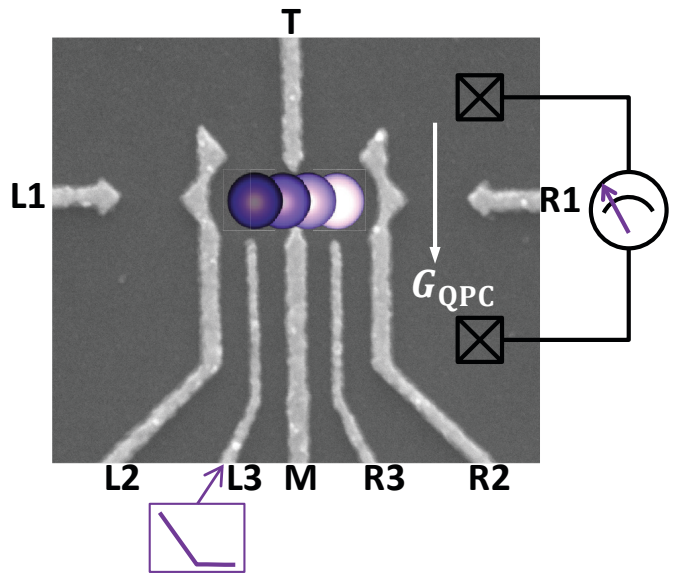

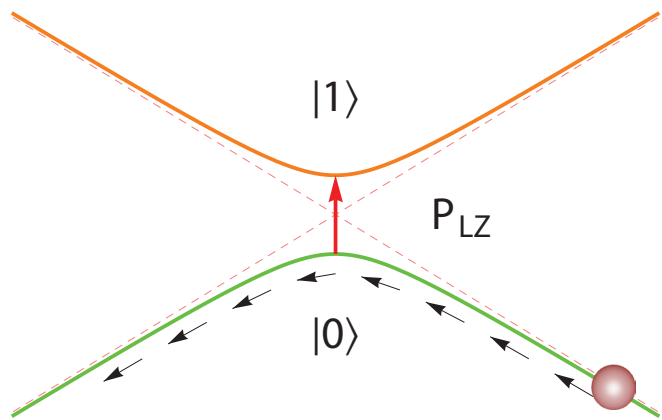 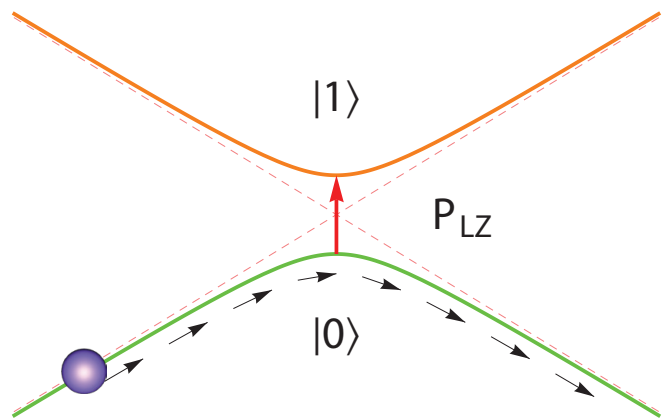

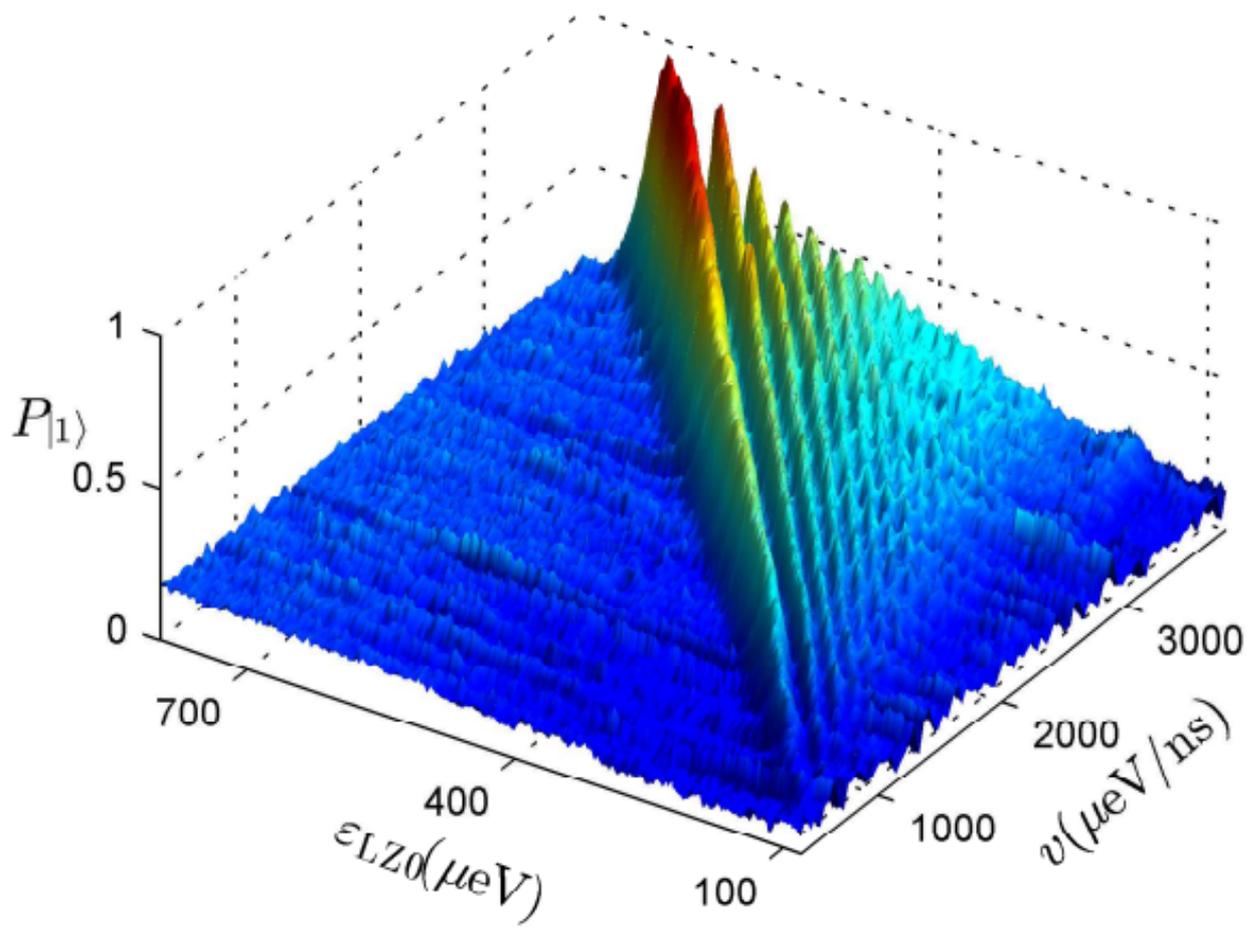

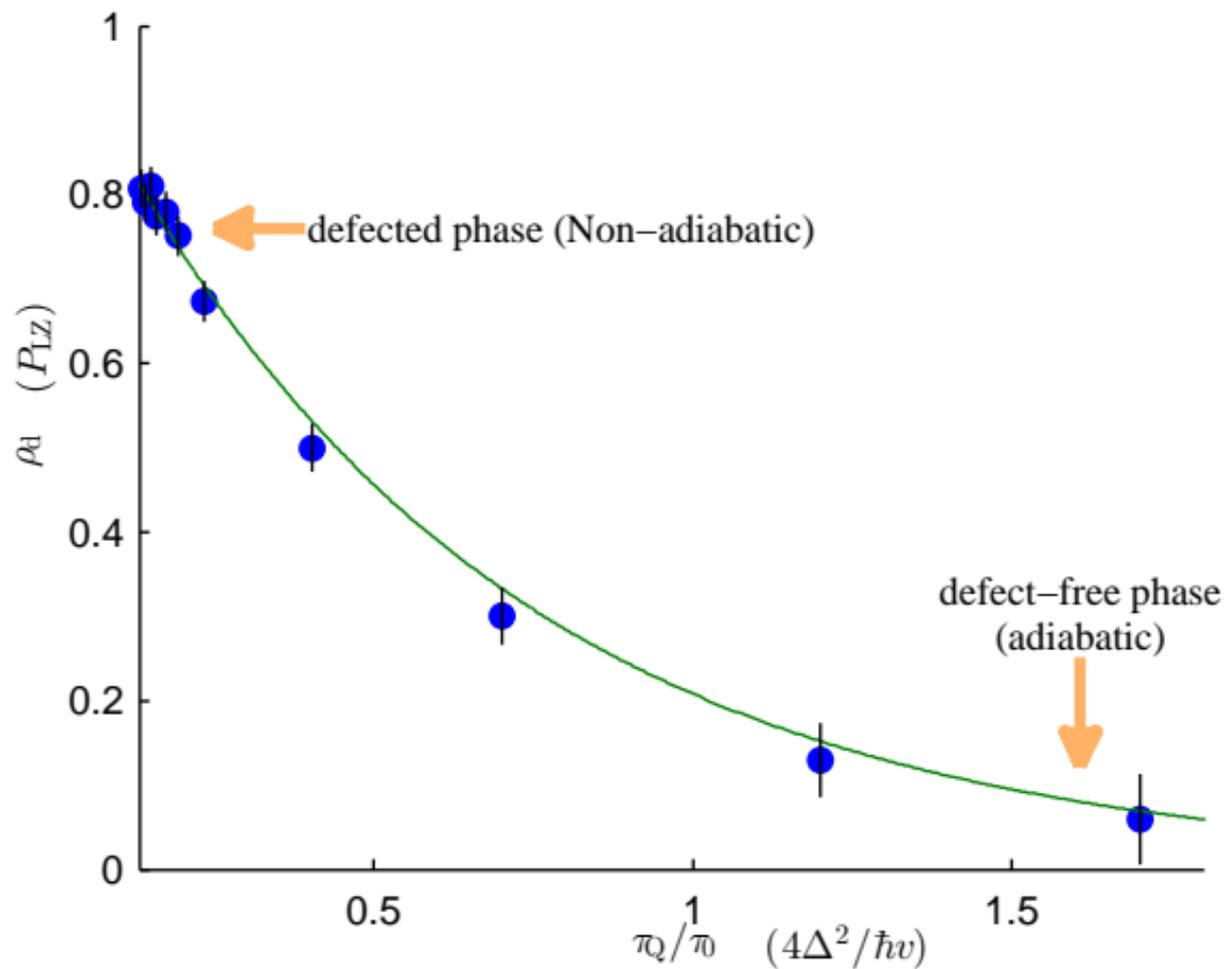